\documentclass[prl,twocolumn,amsmath,amssymb,amsbsy,superscriptaddress,floatfix,nobalancelastpage]{revtex4-1}

\usepackage{graphicx,bm,times}
\usepackage{color}

\begin{document}

\title{Fermi surface of IrTe$_2$ in the valence-bond state as determined by quantum oscillations}

\author{S. F. Blake}
\affiliation{Clarendon Laboratory, Department of Physics, University of Oxford, Parks Road, Oxford OX1 3PU, U.K.}

\author{M. D. Watson}
\affiliation{Clarendon Laboratory, Department of Physics, University of Oxford, Parks Road, Oxford OX1 3PU, U.K.}

\author{A. McCollam}
\affiliation{High Field Magnet Laboratory, Institute
for Molecules and Materials, Radboud University, 6525 ED Nijmegen, The Netherlands}

\author{S. Kasahara}
\affiliation{Department of Physics, Kyoto University, Sakyo-ku, Kyoto 606-8501, Japan}

\author{R. D. Johnson}
\affiliation{Clarendon Laboratory, Department of Physics, University of Oxford, Parks Road, Oxford OX1 3PU, U.K.}

\author{A. Narayanan}
\affiliation{Clarendon Laboratory, Department of Physics, University of Oxford, Parks Road, Oxford OX1 3PU, U.K.}

\author{G. L. Pascut}
\affiliation{Department of Physics and Astronomy, Rutgers University, Piscataway, NJ 08854, USA}

\author{K. Haule}
\affiliation{Department of Physics and Astronomy, Rutgers University, Piscataway, NJ 08854, USA}

\author{V. Kiryukhin}
\affiliation{Department of Physics and Astronomy, Rutgers University, Piscataway, NJ 08854, USA}

\author{T. Yamashita}
\affiliation{Department of Physics, Kyoto University, Sakyo-ku, Kyoto 606-8501, Japan}

\author{D. Watanabe}
\affiliation{Department of Physics, Kyoto University, Sakyo-ku, Kyoto 606-8501, Japan}

\author{T. Shibauchi}
\affiliation{Department of Physics, Kyoto University, Sakyo-ku, Kyoto 606-8501, Japan}

\author{Y. Matsuda}
\affiliation{Department of Physics, Kyoto University, Sakyo-ku, Kyoto 606-8501, Japan}

\author{A. I. Coldea}
\email[corresponding author:]{amalia.coldea@physics.ox.ac.uk}
\affiliation{Clarendon Laboratory, Department of Physics, University of Oxford, Parks Road, Oxford OX1 3PU, U.K.}

\begin{abstract}
We report the observation of the de Haas-van Alphen effect in IrTe$_2$ measured using torque magnetometry at low temperatures down to 0.4~K and in high magnetic fields up to 33~T.
IrTe$_2$ undergoes a major structural transition
around  $\sim 283(1) $~K  due to the formation of planes of Ir and Te dimers
  that cut diagonally through the lattice planes,
  with its electronic structure predicted
 to change significantly from a layered system
 with predominantly three-dimensional character
 to a tilted quasi-two dimensional Fermi surface.
 Quantum oscillations provide direct confirmation of this
 unusual tilted Fermi surface
 and also reveal very light quasiparticle masses (less than 1~$m_e$), with no
 significant enhancement due
 to electronic correlations.
 We find good agreement between the angular dependence of the observed and calculated de Haas-van Alphen frequencies, taking into account the contribution of different structural domains that form while cooling IrTe$_2$.
\end{abstract}

\date{\today}
\maketitle

 Transition metal dichalcogenides are layered metallic compounds which have rich electronic properties.
They often exhibit signatures of unusual electronic behaviour, such as charge density wave (CDW)  \cite{Rossnagel2011}
and superconductivity \cite{Morosan2006,Kiss2007,CastroNeto2001}.
The CDW instability has often been explained as caused
by a partial nesting of the Fermi surface
 \cite{Rossnagel2011,Inosov2008} or, alternatively,
the direct effect of lattice distortions with unexpected local modulations \cite{Dai2014}.
Recently, it has been found that IrTe$_2$ shows a structural transition at
$T_s \sim $280~K~\cite{Fang2013},  which can be suppressed by doping
(for example, Pt or Pd) %\cite{}, Cu~\cite{Kamitani2012} and Rh \cite{Kudo2013,Lei2013};
and a superconducting state is found with a maximum $T_c$ close to
3~K \cite{Ootsuki2012a,Pyon2012,Yang2012}.
There have been numerous scenarios proposed
to explain the nature of the structural transition in IrTe$_2$
linked to the possible Ir $5d$ charge modulation \cite{Yang2012,Kudo2013,Matsumoto1999},
 anionic depolymerisation \cite{Oh2013}, crystal field effects \cite{Fang2013}, the disappearance
of a van Hove singularity close to the Fermi level \cite{Qian2013}
or, similar to other CDW dichalcogenides, associated with the partial Fermi surface nesting  \cite{Ootsuki2012a,Ootsuki2012}.
 However, both optical spectroscopy \cite{Fang2013} and
 angle resolved photoemission spectroscopy (ARPES) \cite{Ootsuki2012,Qian2013} find no evidence of the expected energy gap, and NMR spectra see no CDW behaviour, instead suggesting a lattice-driven transition \cite{Mizuno2002}.
  Furthermore, under applied pressure $T_s$ increases whereas $T_c$ decreases
  \cite{Kiswandhi2013}, opposite to what is expected for a CDW system \cite{CastroNeto2001}.

An alternative scenario for the significant electronic, magnetic
and structural changes at $T_s$ in IrTe$_2$ is based on  detailed
understanding of the lattice effects as a function of temperature \cite{Pascut}.
These studies suggest that the low temperature (LT) phase of IrTe$_2$ is an unexpected
{\it valence-bond state} (VBS)~\cite{Pascut}
with a highly non-sinusoidal structural modulation with fundamental vector
${\bf q_0}=( \frac{1}{5}, 0, \frac{1}{5})$
\cite{Yang2012,Oh2013,Cao2013a,Hsu2013}.
The predicted electronic  structure  of  the  low  temperature  modulated  phase  is  strongly  influenced  by    Ir  (and,  to  a lesser extent, Te) dimerization. These dimers form planes that cut diagonally through the structural IrTe$_6$ octahedral planes, with  reduced density of states at the Fermi level (see Fig.~1a).
 This dimerization is predicted to cause a significant change in the electronic
  structure, from a predominantly quasi-three-dimensional (Q3D) Fermi surface in the high temperature phase (HT) to a highly unusual quasi-two dimensional (Q2D) Fermi surface
tilted away from the lattice planes
in the low temperature (LT) phase below $T_s$ \cite{Pascut,Toriyama2014}. This LT phase is
a unique electronic structure among metal dichalocogenides.

Experimental knowledge of the Fermi surface is vital
if we are to conclusively address these predictions and
understand the novel phenomena of valence-bond formation in IrTe$_2$.
ARPES  measurements  of  IrTe$_2$  at  room  temperature  consistently  find  agreement  with  the  Fermi surfaces predicted from first-principle calculations \cite{Ootsuki2012,Qian2013}, but fail to provide a clear picture of the changes that occur below $T_s$. Two of the challenges which are likely to impede these experiments significantly are:  i) the unusual  tilting  of  the  proposed  Q2D  Fermi  surface  away  from  the  cleaving  plane \cite{Ootsuki2013,Ootsuki2013a};  ii)  domain formation  due  to  the  lowering  in  symmetry through  the  structural  transition.
In this Letter we present a de Haas-van Alphen (dHvA) study
in the valence-bond state of IrTe$_2$,
which provides direct experimental evidence
for the unusual tilted Q2D Fermi surface.
 The experimentally measured effective masses are relatively light, and do not show
 any significant enhancement above the
 band mass, suggesting that electronic correlations
 do not play a major role in this metallic system.
 We find that in order to fully account for the observed data, the domain formation
through the structural transition must be considered.

\begin{figure}[h!]
\centering
    \includegraphics[width=9cm,clip=true]{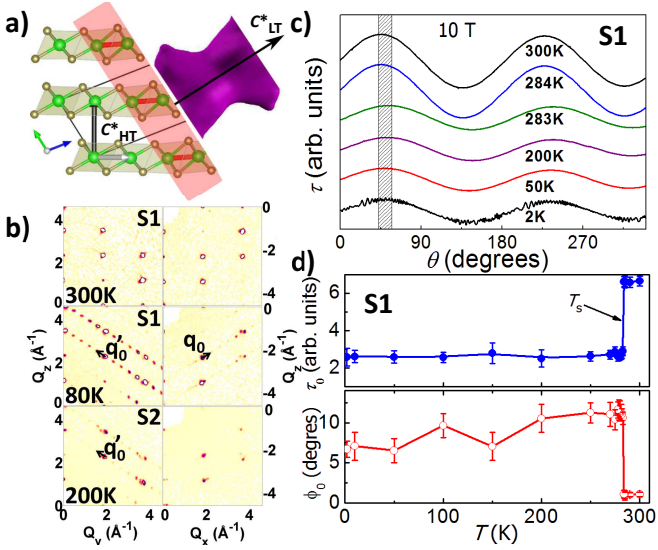}
  \caption{ The effect of the structural transition in IrTe$_2$.
  a) The HT and LT reciprocal vectors, $c^*_{HT}$ and $c^*_{LT}$, are indicated together with a representation of a tilted Q2D Fermi surface in the VBS state.
  b) Single crystal X-ray diffraction taken above and below $T_s\sim 283.5$~K. Below $T_s$ a superstructure forms with vectors ${\bf q}'_0$=(0,−1/5,1/5) and ${\bf q}_0$=(1/5, 0, 1/5) within the $0kl$ and $h0l$ planes for sample S1 (with two domains) and and only ${\bf q}'_0$  for sample S2 (one domain), all indexed in the HT hexagonal unit cell.
  c) Angular dependence of torque for sample S1 at constant magnetic field (10~T)
  and temperatures. The shaded region shows the shift in angle while cooling through $T_s$.
     d) The temperature dependence of the torque signal, $\tau_0$ and the shift in the phase, $\phi_0$,
     (as defined in the text) showing strong anomalies at $T_s$.
The rotation in the ($ac$) plane is always defined
in the HT rhombohedral structure ($P\bar{3}m1$ symmetry).}
\label{char}
\end{figure}
IrTe$_2$ crystals were grown using Te flux, as reported previously \cite{Pyon2013}.
X-ray data were taken on different single crystals as a function of temperature,
using an Agilent SuperNova Diffractometer,
fitted with liquid nitrogen open-flow cooling (Oxford Cryosystems).
The torque measurements were carried out
using piezo-resistive, self sensing cantilevers,
on more than four
single crystals with sizes usually smaller
than $70 \times 70 \times 10 ~\mu$m$^3$ (also characterized using X-rays)
down to 0.3~K, in static magnetic fields up to
18~T in Oxford and 33~T at the HFML in Nijmegen.
Throughout this paper, $\theta$ is
defined as the angle between the magnetic field direction and
the crystallographic $c$-axis in the HT rhombohedral structure
 (spacegroup $P \bar{3} m1$) within the $(ac)$ plane,
 such that $\theta=0^\circ$
 corresponds to ${\bm H}\| [001]$ (along $c_{\rm HT}$ axis),
  $\theta=90^\circ$ to ${\bm H}\| [100]$ (along $a_{\rm HT}$ axis)
 and $\theta=-90^\circ$ to ${\bm H}\| [\bar{1}00]$.
 We calculated the band-structure
 using the linear augmented plane wave method, with generalised gradient approximations
implemented in WIEN2k \cite{Blaha2001}
using the structural parameters
from Ref.\cite{Pascut}
\footnote{The cutoff for plane
wave expansion in the interstitials was $R \times K_{max}=9$
(convergence is reached for $R \times K_{max}=8$) and 5472 k-points
were used in the first Brillouin zone.}.
Spin-orbit coupling was included for all calculations.
The extremal Fermi surface areas for the triclinic $P\bar{1}$ symmetry
were calculated using SKEAF \cite{Rourke2012}.

Figure \ref{char}b) (top panels) shows two typical sets of diffraction patterns above  $T_s$,
for two different samples S1 and S2, indexed using a HT hexagonal unit cell.
The diffraction patterns measured below  $T_s$,  show
additional superstructure peaks at ${\bf q}'_0$=(0,−1/5,1/5) and ${\bf q}_0$=(1/5, 0, 1/5) [see Figure 1b) middle and bottom panels]. These extra reflections are visible between the main diffraction spots of the HT unit cell, in both the $0kl$ and $h0l$ planes for sample S1, indicating the presence two domains, but they are present in only one direction for sample S2, suggesting that it contains a single domain \cite{Oh2013,Pascut}.
 STM measurements found that different crystals of IrTe$_2$ could
display additional charge modulations,
${\bf q_n}$=$(3n+2)^{-1}$,
over a large temperature range \cite{Hsu2013}.
However our low temperature  X-ray measurements, performed well
below $T_s$, show only the presence of 1/5 modulation in  both
samples S1 and S2
(as also found in Ref.\cite{Toriyama2014}),
and we find further evidence of domain
formation in our torque measurements.

 Magnetic torque is caused by magnetic susceptibility anisotropy,
measuring the misalignment of the magnetization with respect to a
uniform applied field, ${\bm H}$.
It is given by ${\bm \tau}={\bm M}  \times \mu_0 {\bm H}$ where ${\bm M}$
is the bulk magnetization.
For IrTe$_2$, if ${\bm M}$ and ${\bm H}$ lie in the ($ac$) plane, then $\tau=
\mu_0 (M_a H_c- M_c H_a ) = \tfrac{1}{2} \mu_0 \Delta \chi_{ac}
H^2$ $\sin (2\theta$-$\phi_0$)= $\tau_0 \sin (2\theta$-$\phi_0$),
where $\Delta \chi_{ac}$ is the susceptibility anisotropy between the
(HT rhombohedral) $c$-axis and $a$-axis
 and $\phi_0$ is the angle shift that arises
 due to change in symmetry and tilting of the $c^*$ axis
relative to the HT structure [see also Fig.1a)].
 Fig.~\ref{char}c) shows the angular dependence of torque for a single crystal of IrTe$_2$  (S1),
 at different temperatures, in a constant applied magnetic field of 10~T.
Using the above expression for $\tau$,
we fitted the data in Fig.~\ref{char}d) for sample S1,
and found that
there is large reduction of amplitude, $\Delta \tau_0 \approx$ 40 \%,
through the structural transition at $T_s$
(proportional to $\Delta \chi_{ac}$)
and a significant  phase shift.
The decrease in amplitude at $T_s$ is
likely to be caused by the change
in susceptibility anisotropy as the system lowers it symmetry,
but the change in electronic susceptibility  \cite{Matsumoto1999},
caused by a decrease in the density of states
at the Fermi level \cite{Fang2013} could also play a role.
 We observe an
 angular shift of $\phi_0 \sim 10^\circ$, while cooling through $T_s$, [Fig.\ref{char}d)],
 which is likely to be associated with the tilting of the $c^{*}$ axis below $T_s$
 such that it is no longer perpendicular to the lattice planes,
as expected in the VBS
(shown schematically in Fig.\ref{char}a) \cite{Pascut,Toriyama2014}).

\begin{figure}[h!]
\centering
    \includegraphics[width=7cm]{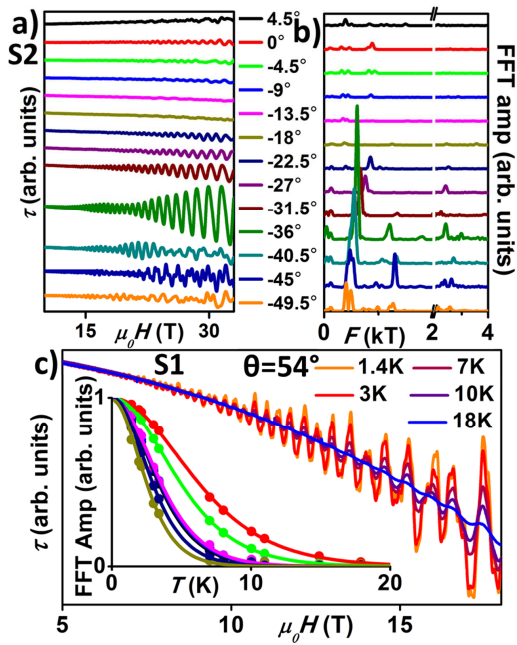}
  \caption{Quantum oscillations in IrTe$_2$. a) Field dependence of torque for sample S2 measured at 0.4~K for different orientations in magnetic field.  b) The corresponding FFT spectra (with amplitudes scaled as $A^{3/4}$) using a large field window $5-33$ T to separate lower frequencies below 2~kT
  and $20-33$~T to detect higher frequencies. c) Temperature dependence of
  quantum oscillations for sample S1 at $\theta=54^{\circ}$. The inset shows
   the temperature dependence of the scaled FFT amplitudes for individual frequencies
   fitted to the Lifshitz-Kosevich formula (8-18~T) to extract the effective masses, $m^*$ \cite{Lifshitz1956}.}
\label{raw}
\end{figure}

Torque magnetometry in metallic systems at low temperatures
allows access to anisotropic Fermi surfaces through
the observation of quantum oscillations (de Haas-van Alphen effect)
and has been successfully used to measure
the Fermi surface of other dichalocgenides, such as 2H-NbSe$_2$ \cite{Steep1995}.
Fig.\ref{char}c) shows clear quantum oscillations at 2~K
superimposed on a background signal, as measured for sample S1.
In Fig.~\ref{raw}a) we show the field dependence of torque for sample S2
measured at 1.4~K for different
orientations in magnetic field. Quantum oscillations
are clearly observed in the raw torque signal
over a wide angular range in different single crystals of IrTe$_2$.
The oscillation frequencies were extracted
using a fast Fourier transform (FFT) of the raw data
(after subtracting a low order polynomial to correct for background torque).
Each of these frequencies $F_i$ are related
to the extremal cross-section areas $A_i$ of the Fermi surface
by the Onsager relation, $F_i = \hbar A_i/(2 \pi e)$, for a particular
field orientation, as shown in Fig.\ref{raw}b).
The spectra close to $\theta=0^{\circ}$ are dominated by multiple frequencies
below 1~kT that are observed over a large angular range.
In high magnetic fields (above 15~T) we can also detect the
presence of additional, higher frequencies around 1.5--4~kT,
but no further  frequencies were observed in fields up to
33~T and temperatures down to 0.3~K.

Fig.~\ref{raw}c) shows the temperature dependence of the
quantum oscillations for sample S1 that are visible up to 20~K
for  $\theta=54^{\circ}$.
The inset shows the temperature dependence of the
corresponding FFT amplitudes fitted to the Lifshitz-Kosevich expression \cite{Lifshitz1956}
in order to extract the effective masses corresponding to each individual frequency.
The extracted values of the effective masses for different orientations
in magnetic field are listed in Table~\ref{table}.
We find similar values of frequencies and masses for different samples when the magnetic field
is aligned along the $c$ axis in the HT phase.
 All masses were small, below $1.1 m_e$, suggesting that they originate
 from rather wide $5d$ and $5p$ bands with Ir and Te character \cite{Pascut,Fang2013}.

\begin{figure*}[htbp]
\centering
\includegraphics[height=7.0cm,clip=true]{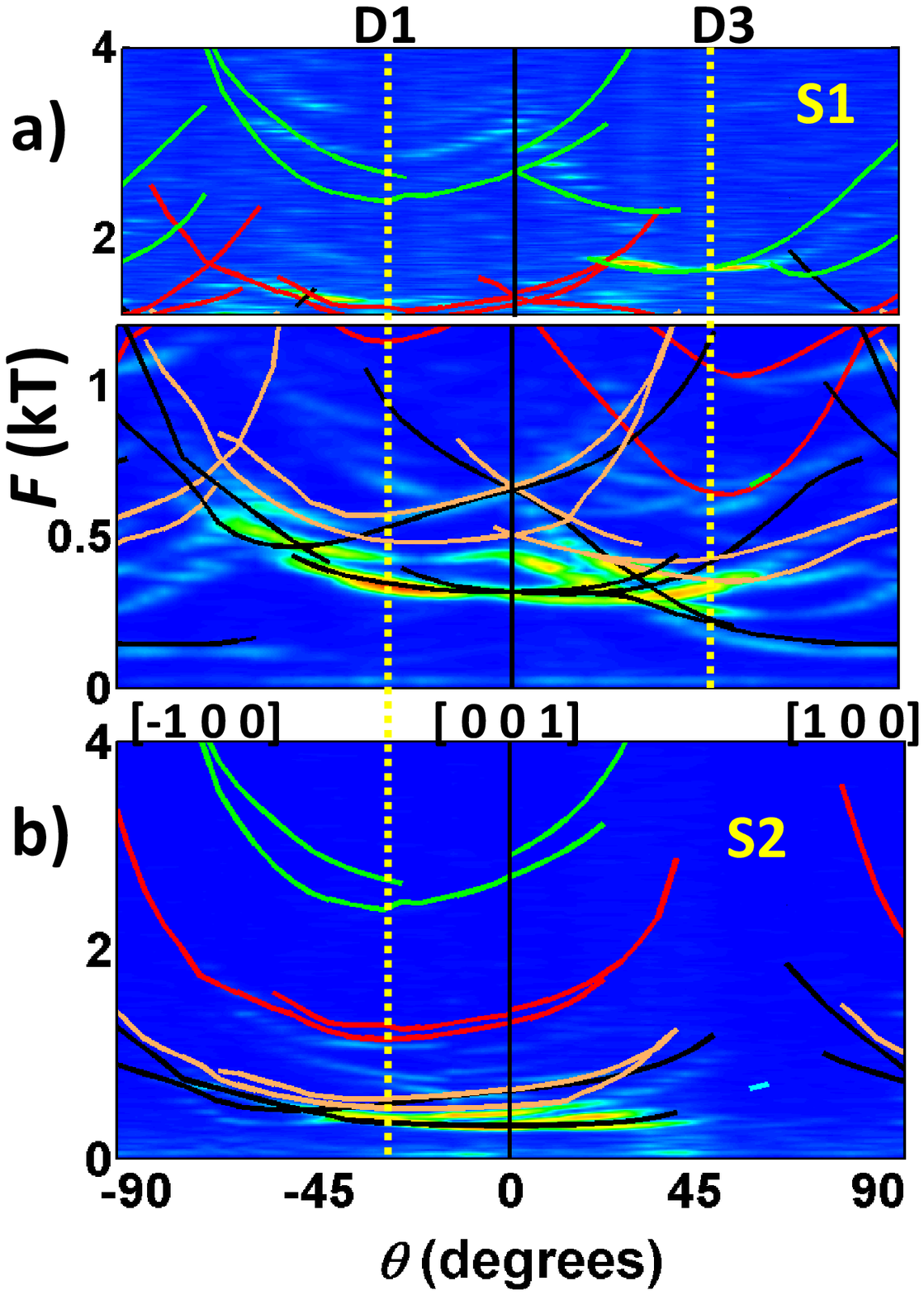}
\includegraphics[height=7.0cm,clip=true]{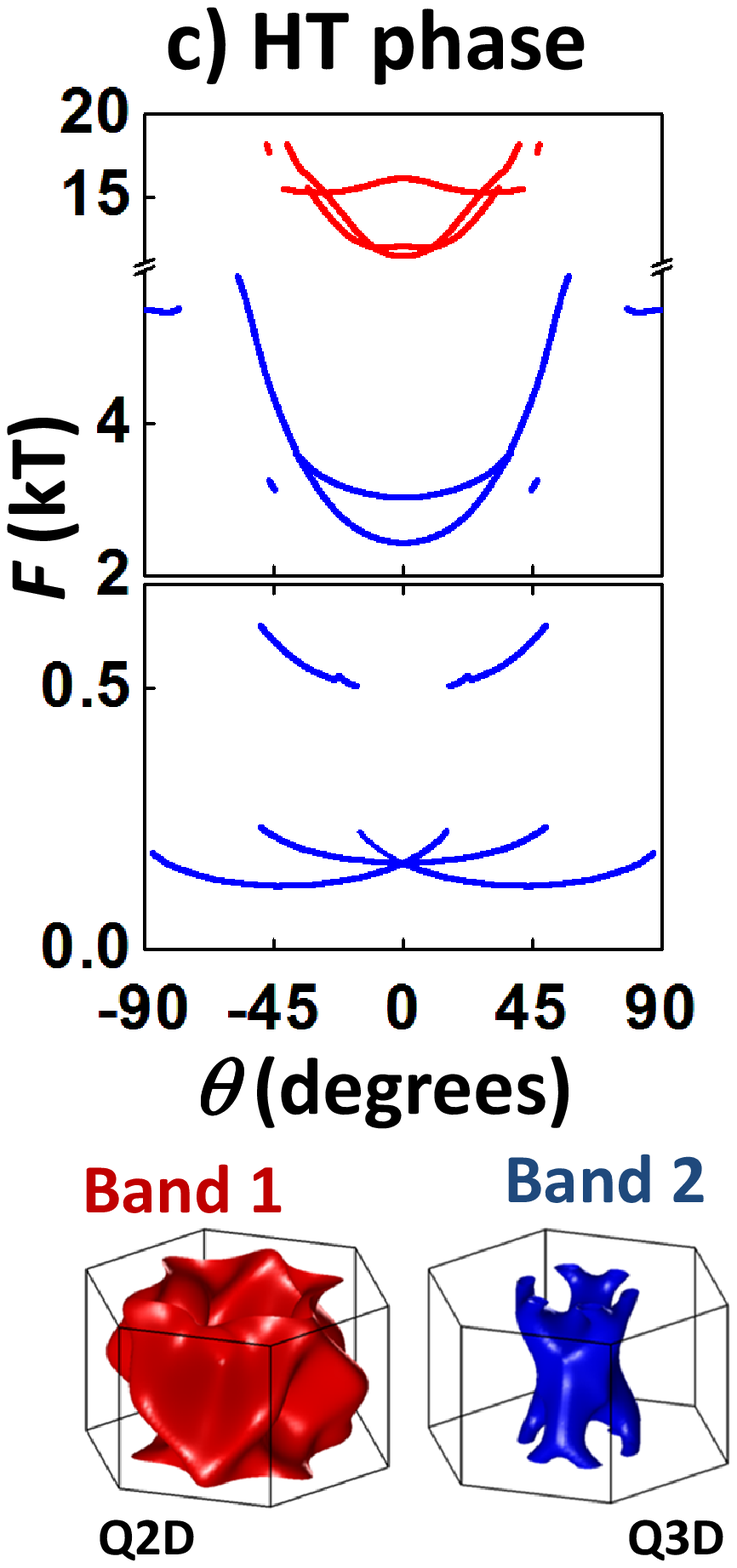}
\includegraphics[height=7.0cm,clip=true]{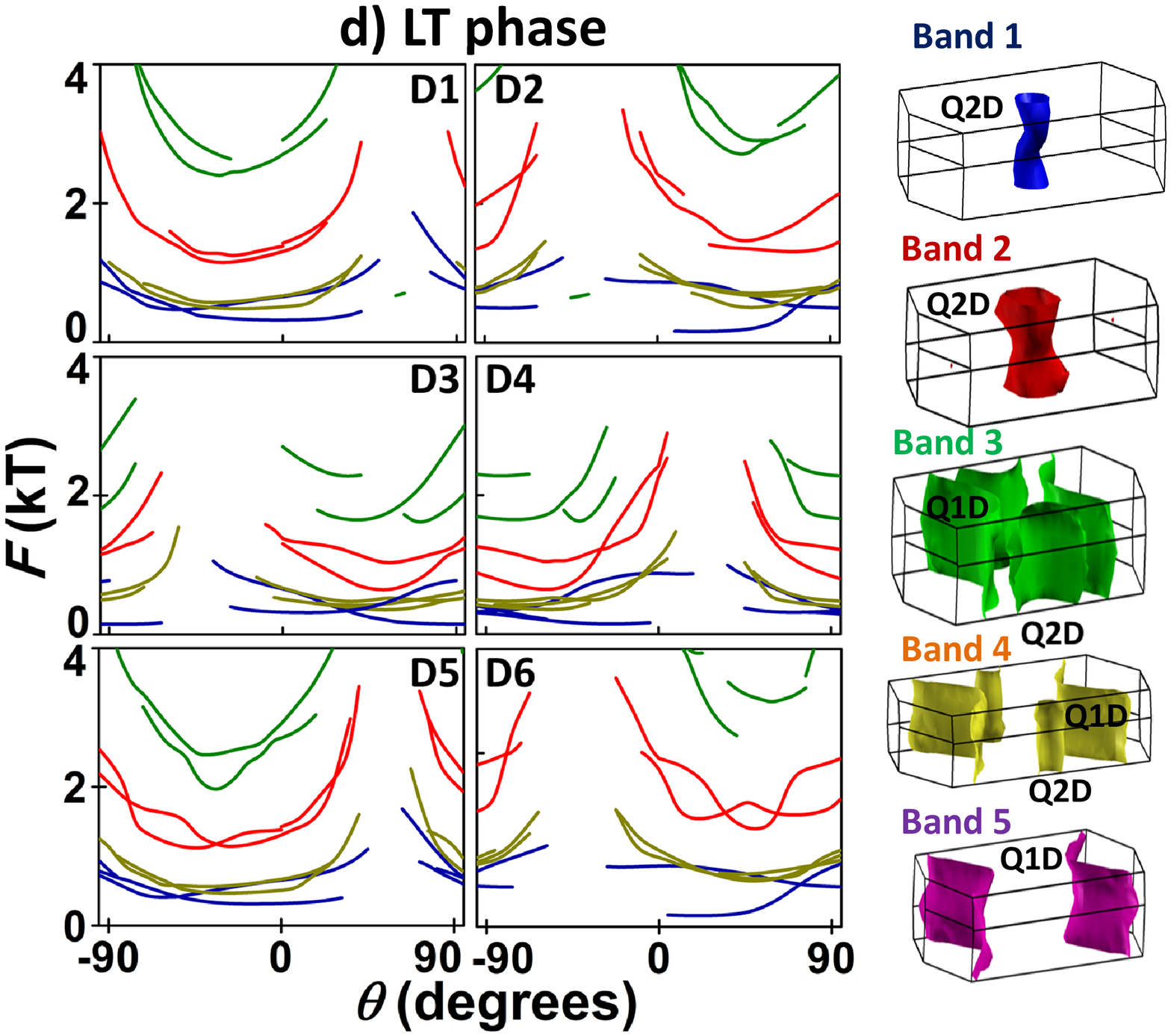}
   \caption{Comparison between the measured and calculated angular dependence of the dHvA frequencies of IrTe$_2$  for a) sample S1 (top panels) and b) S2 (bottom panel). The experimental data are compared with calculations (solid lines) for domains D1 and D3 for sample S1 and domain D1 for sample S2, with the local dHvA frequency minima for each domain indicated by the dashed lines. In the colourmaps, the FFT amplitudes are scaled as $A^{3/4}$ and the data are collected every 4.5 degrees. The frequency branches not compared to simulations correspond to the second harmonics of the lower, stronger frequencies. The calculated angular dependence of the predicted dHvA frequencies and the Fermi surfaces corresponding to the c) HT structure and d) LT dimerized structure for the six possible domains D1-6. The Brillouin zones of both structures are indicated by the solid lines.}
\label{Fig3}
\end{figure*}

In order to be able to determine the shape of the Fermi surface
we need to understand the overall angular dependence of the
observed frequencies in IrTe$_2$.
Figs.~\ref{Fig3}a) and b) show
detailed angular dependence of the observed dHvA frequencies
plotted as a colour map of the FFT amplitude
for samples S1  and S2, respectively.
We observe that the frequencies
are periodic over $180^\circ$, but have no further symmetry.
In the case of a Q2D
 Fermi surface, one would expect frequency branches with minima
 at $\theta=0^{\circ}$ and an $F / \cos \theta$ dependence,
 which would disappear before
 reaching $\theta=90^{\circ}$ [see Fig.~\ref{Fig3}c)].
We have calculated the expected frequencies for the HT
 rhombohedral Q3D Fermi surface that is composed of a strongly corrugated
cylindrical Fermi surface
dominated by the Ir $5d$ orbitals [band 1 in Fig.~\ref{Fig3}c)] and
 a complex inner band of predominantly Te $5p$ orbital character
 (band 2 in Fig.~\ref{Fig3}c).
 The outer band 1 produces frequencies in excess of 10~kT [see Fig.~\ref{Fig3}c)] and rather
heavy masses (1.5 to 1.8 $m_e$) that are much larger than those observed experimentally.
However, the inner band shows a mixture
of Q2D Fermi surfaces with frequencies below 5~kT,
and  Q3D ones  with
frequencies below 1~kT that extend over the entire angular
range, shown in Fig.~\ref{Fig3}c).
The calculated frequencies and masses for the inner band (band 2) are in a similar range
to those we observe experimentally, but the number of frequencies,
their angular dependence and the lack of symmetry around $\theta=0^\circ$
 suggest that the Fermi surface of IrTe$_2$ in the LT phase is quite
 different to that of the HT phase, and originates
from a lower symmetry structure.
Moreover, our frequency spectra have minima away from the rhombohedral axes, either close to
$\theta \approx -30^{\circ}$ or $\theta \approx 54^{\circ}$ for S1 [indicated by
  the dashed lines in Fig.~\ref{Fig3}a) and b)], suggesting that the $c^{*}$  in the LT phase axis is tilted relative to that of the HT phase.
 The most likely reason for this behaviour is the
unusual electronic structure resulting from the $1/5$-modulated triclinic structure \cite{Pascut} (Fig.~\ref{char}a)), also suggested by our X-ray diffraction data on the same crystals [Fig.~\ref{char}b)].
The predicted Fermi surface in the LT phase is shown in Fig.\ref{Fig3}d),
and is composed of multiple Q2D bands
as well as Q1D bands (the latter would not give any
quantum oscillation signal due to the lack of closed extremal orbits for any orientation
of the magnetic field).
As the symmetry axis for the quasi-two dimensional bands is tilted away
from the HT $c_{\rm HT}^*$ direction,
we would expect the location of the minima in frequency  to be significantly shifted, as we see in experiments.

While cooling through the phase transition at $T_s$, single crystals of IrTe$_2$
are prone to domain formation, as the direction of dimerisation must be chosen
 while changing symmetry \cite{Pascut}.
 The $P\bar{3}m1$ rhombohedral space group has twelve symmetry operations,
 while the triclinic $P\bar{1}$ has two;
upon cooling from $P\bar{3}m1$ to $P\bar{1}$  six possible domains can form
 [see Supplementary Material (SM)].
Using the bandstructure calculations for the dimerized state
of IrTe$_2$ \cite{Pascut}, we have calculated the predicted
dHvA frequencies for the different domains, as shown in Fig.~\ref{Fig3}d).
We find that the simulated spectra of frequencies
show local minima in different locations for different domains,
and this helps us
to identify which domains are present in our crystals.
In Fig.~\ref{Fig3}a) we compare the measured frequencies for sample S1
with those calculated for each domain,
and a good fit is found by considering two domains [domain D1 with local minima
in frequency at $\theta \sim -30^{\circ} $ and domain D3 with local minima at $\theta \sim 54^{\circ}$,
 indicated by the dotted lines in Fig.\ref{Fig3}a)].
The torque data for Sample S2 can be described by the presence of a single domain, domain D1,
  as shown in Fig.\ref{Fig3}b), in good agreement with the X-ray data shown in Fig.~\ref{char}b).
However, in sample S1 the two domains revealed by the torque measurements were
also present in the low temperature x-ray measurements [Fig.~\ref{char}b)].

Comparisons of calculated frequencies and effective masses with our experimental data for different orientations are listed in Table~\ref{table}. The calculated values of the frequencies and masses are in the same range as those measured for sample S1. We find that the smallest measured orbit of 120~T
matches well with the smallest possible calculated orbit of 190~T in the $P\bar{1}$ structure and
there is overall good agreement between the measured and calculated values.
No significant mass enhancement above the band mass $m_b$
is detected, as shown in Table~\ref{table}.
We can account for all four Q2D sheets of the Fermi surface
in the LT phase, but we notice that
the measured frequencies of band 4 are consistently $\sim 100$ T smaller than simulations,
suggesting that this Q2D cylinder is slightly
smaller than those predicted by calculations; this effect may
be due to small variation in the lattice parameters, as calculations are
considering a LT structure at 220~K \cite{Pascut} and further variations
in amplitudes for sample S2 as compared with sample S1 are
due to the smaller crystal size
\footnote{For large $F$ and $m^*$,
the exponential damping of the dHvA signal due impurity scattering will hamper
the observation of those dHvA oscillations, as compared with small $F$
and $m^*$. Any small misalignment of the samples could cause
small shifts in the positions of the dHvA frequencies}.
The presence of a quasi-two dimensional Fermi surface
tilted away from the main cleaving planes is likely to have a
detrimental effect on surface sensitive probes, such as ARPES,
where the underlying Fermi surface of IrTe$_2$
in the VBS would be obscured
due to the strong $k_z$
dispersion~\cite{Qian2013}.

\begin{table}[h]
\caption{Comparison between the measured quantum oscillations
frequencies, $F$, and effective masses, $m^*$,
for sample S1 and the calculated
values ($m_b$, band mass) for the LT dimerized state of IrTe$_2$
in the presence of two domains (D1 and D3) and
for two $\theta$ orientations in magnetic field.}
\begin{center}
\begin{tabular}{c c c c c c c c c}
\hline
\hline
 & \multicolumn{4}{c}{ $\theta=0^\circ$} & \multicolumn{4}{c}{ $\theta=54^\circ$}  \\
 & \multicolumn{2}{c}{calculations} & \multicolumn{2}{c}{experiment} & \multicolumn{2}{c}{calculations} & \multicolumn{2}{c}{experiment}\\
\hline
 & $F$ & $m_b$ & $F$ & $m^*$ & $F$ & $m_b$ & $F$ & $m^*$\\
 & (T) & ($m_e$) & (T) & ($m_e$) & (T) & ($m_e$) & (T) & ($m_e$)\\
\hline
1a & 318 & 0.27 & 303(1) & 0.34(1) & 194 & 0.33 & 164(2) & 0.50(1)\\
1b & 655 & 0.59 & 689(7) & 0.56(1) & 461 & 0.34 & 357(1) & 0.34(1)\\
4a & 508 & 0.29 & 420(1) & 0.31(1) & 354 & 0.27 & 246(9) & 0.27(1)\\
4b & 667 & 0.42 & 588(3) & 0.47(1) & 523 & 0.38 & - & -\\
2a & 1305 & 0.73 & 1266(4) & 0.72(5) & 640 & 0.55 & 607(1) & 0.48(1)\\
2b & 1426 & 0.88 & 1530(10) & 0.89(2) & 1032 & 0.58 & 1024(6) & 0.66(2)\\
3a & 2705 & 0.77 & - & - & 658 & 0.58 & 674(10) & 0.47(1)\\
3b & 2911 & 1.20 & 2968(8) & 0.98(1) & 1742 & 0.53 & 1688(8) & 0.57(1)\\
\hline
\hline
\end{tabular}
\end{center}
\label{table}
\end{table}

To summarize, we have
experimentally determined the
Fermi surface of IrTe$_2$ in the valence-bond state
as being quasi-two dimensional with the
symmetry axis tilted away from that of the high temperature layered structure,
in good agreement with the first-principles band structure calculations.
 The measured effective masses are very light and close to the
 the calculated band masses, suggesting that electronic correlations do not
 play an important role in IrTe$_2$.
 To fully describe our data the contribution of different domains,
 which form through the structural
 phase transition, needs to be considered.
These results confirm that below the structural transition IrTe$_2$
becomes a weakly correlated metal with
 a tilted quasi-two dimensional electronic structure
 in its valence-bond state.

We thank Robert Schoonmaker for computational support.
This work was supported by EPSRC
(EP/L001772/1, EP/I004475/1, EP/I017836/1) and
part of the work was performed at the HFML/RU-FOM, member of the
European Magnetic Field Laboratory (EMFL). AIC
acknowledges an EPSRC Career Acceleration Fellowship.
G.L.P. was supported by the
NSF under Grant No. DMR-1004568, and
V.K. by NSF DMREF Grant No. 12-33349.

\bibliography{IrTe2paper_june14}

\begin{thebibliography}{30}%
\makeatletter
\providecommand \@ifxundefined [1]{%
 \@ifx{#1\undefined}
}%
\providecommand \@ifnum [1]{%
 \ifnum #1\expandafter \@firstoftwo
 \else \expandafter \@secondoftwo
 \fi
}%
\providecommand \@ifx [1]{%
 \ifx #1\expandafter \@firstoftwo
 \else \expandafter \@secondoftwo
 \fi
}%
\providecommand \natexlab [1]{#1}%
\providecommand \enquote  [1]{``#1''}%
\providecommand \bibnamefont  [1]{#1}%
\providecommand \bibfnamefont [1]{#1}%
\providecommand \citenamefont [1]{#1}%
\providecommand \href@noop [0]{\@secondoftwo}%
\providecommand \href [0]{\begingroup \@sanitize@url \@href}%
\providecommand \@href[1]{\@@startlink{#1}\@@href}%
\providecommand \@@href[1]{\endgroup#1\@@endlink}%
\providecommand \@sanitize@url [0]{\catcode `\\12\catcode `\$12\catcode
  `\&12\catcode `\#12\catcode `\^12\catcode `\_12\catcode `\%12\relax}%
\providecommand \@@startlink[1]{}%
\providecommand \@@endlink[0]{}%
\providecommand \url  [0]{\begingroup\@sanitize@url \@url }%
\providecommand \@url [1]{\endgroup\@href {#1}{\urlprefix }}%
\providecommand \urlprefix  [0]{URL }%
\providecommand \Eprint [0]{\href }%
\providecommand \doibase [0]{http://dx.doi.org/}%
\providecommand \selectlanguage [0]{\@gobble}%
\providecommand \bibinfo  [0]{\@secondoftwo}%
\providecommand \bibfield  [0]{\@secondoftwo}%
\providecommand \translation [1]{[#1]}%
\providecommand \BibitemOpen [0]{}%
\providecommand \bibitemStop [0]{}%
\providecommand \bibitemNoStop [0]{.\EOS\space}%
\providecommand \EOS [0]{\spacefactor3000\relax}%
\providecommand \BibitemShut  [1]{\csname bibitem#1\endcsname}%
\let\auto@bib@innerbib\@empty
%</preamble>
\bibitem [{\citenamefont {Rossnagel}(2011)}]{Rossnagel2011}%
  \BibitemOpen
  \bibfield  {author} {\bibinfo {author} {\bibfnamefont {K.}~\bibnamefont
  {Rossnagel}},\ }\href@noop {} {\bibfield  {journal} {\bibinfo  {journal}
  {Journal of Physics: Condensed Matter}\ }\textbf {\bibinfo {volume} {23}},\
  \bibinfo {pages} {213001} (\bibinfo {year} {2011})}\BibitemShut {NoStop}%
\bibitem [{\citenamefont {Morosan}\ \emph {et~al.}(2006)\citenamefont
  {Morosan}, \citenamefont {Zandbergen}, \citenamefont {Dennis}, \citenamefont
  {Bos}, \citenamefont {Onose}, \citenamefont {Klimczuk}, \citenamefont
  {Ramirez}, \citenamefont {Ong},\ and\ \citenamefont {Cava}}]{Morosan2006}%
  \BibitemOpen
  \bibfield  {author} {\bibinfo {author} {\bibfnamefont {E.}~\bibnamefont
  {Morosan}}, \bibinfo {author} {\bibfnamefont {H.~W.}\ \bibnamefont
  {Zandbergen}}, \bibinfo {author} {\bibfnamefont {B.~S.}\ \bibnamefont
  {Dennis}}, \bibinfo {author} {\bibfnamefont {J.~W.~G.}\ \bibnamefont {Bos}},
  \bibinfo {author} {\bibfnamefont {Y.}~\bibnamefont {Onose}}, \bibinfo
  {author} {\bibfnamefont {T.}~\bibnamefont {Klimczuk}}, \bibinfo {author}
  {\bibfnamefont {A.~P.}\ \bibnamefont {Ramirez}}, \bibinfo {author}
  {\bibfnamefont {N.~P.}\ \bibnamefont {Ong}}, \ and\ \bibinfo {author}
  {\bibfnamefont {R.~J.}\ \bibnamefont {Cava}},\ }\href@noop {} {\bibfield
  {journal} {\bibinfo  {journal} {Nature Physics}\ }\textbf {\bibinfo {volume}
  {2}},\ \bibinfo {pages} {544} (\bibinfo {year} {2006})}\BibitemShut {NoStop}%
\bibitem [{\citenamefont {Kiss}\ \emph {et~al.}(2007)\citenamefont {Kiss},
  \citenamefont {Yokoya}, \citenamefont {Chainani}, \citenamefont {Shin},
  \citenamefont {Hanaguri}, \citenamefont {Nohara},\ and\ \citenamefont
  {Takagi}}]{Kiss2007}%
  \BibitemOpen
  \bibfield  {author} {\bibinfo {author} {\bibfnamefont {T.}~\bibnamefont
  {Kiss}}, \bibinfo {author} {\bibfnamefont {T.}~\bibnamefont {Yokoya}},
  \bibinfo {author} {\bibfnamefont {A.}~\bibnamefont {Chainani}}, \bibinfo
  {author} {\bibfnamefont {S.}~\bibnamefont {Shin}}, \bibinfo {author}
  {\bibfnamefont {T.}~\bibnamefont {Hanaguri}}, \bibinfo {author}
  {\bibfnamefont {M.}~\bibnamefont {Nohara}}, \ and\ \bibinfo {author}
  {\bibfnamefont {H.}~\bibnamefont {Takagi}},\ }\href@noop {} {\bibfield
  {journal} {\bibinfo  {journal} {Nature Physics}\ }\textbf {\bibinfo {volume}
  {3}},\ \bibinfo {pages} {720} (\bibinfo {year} {2007})}\BibitemShut {NoStop}%
\bibitem [{\citenamefont {{Castro Neto}}(2001)}]{CastroNeto2001}%
  \BibitemOpen
  \bibfield  {author} {\bibinfo {author} {\bibfnamefont {A.}~\bibnamefont
  {{Castro Neto}}},\ }\href@noop {} {\bibfield  {journal} {\bibinfo  {journal}
  {Phys. Rev. Lett.}\ }\textbf {\bibinfo {volume} {86}},\ \bibinfo {pages}
  {4382} (\bibinfo {year} {2001})}\BibitemShut {NoStop}%
\bibitem [{\citenamefont {Inosov}\ \emph {et~al.}(2008)\citenamefont {Inosov},
  \citenamefont {Zabolotnyy}, \citenamefont {Evtushinsky}, \citenamefont
  {Kordyuk}, \citenamefont {Buechner}, \citenamefont {Follath}, \citenamefont
  {Berger},\ and\ \citenamefont {Borisenko}}]{Inosov2008}%
  \BibitemOpen
  \bibfield  {author} {\bibinfo {author} {\bibfnamefont {D.~S.}\ \bibnamefont
  {Inosov}}, \bibinfo {author} {\bibfnamefont {V.~B.}\ \bibnamefont
  {Zabolotnyy}}, \bibinfo {author} {\bibfnamefont {D.~V.}\ \bibnamefont
  {Evtushinsky}}, \bibinfo {author} {\bibfnamefont {A.~A.}\ \bibnamefont
  {Kordyuk}}, \bibinfo {author} {\bibfnamefont {B.}~\bibnamefont {Buechner}},
  \bibinfo {author} {\bibfnamefont {R.}~\bibnamefont {Follath}}, \bibinfo
  {author} {\bibfnamefont {H.}~\bibnamefont {Berger}}, \ and\ \bibinfo {author}
  {\bibfnamefont {S.~V.}\ \bibnamefont {Borisenko}},\ }\href {\doibase
  {10.1088/1367-2630/10/12/125027}} {\bibfield  {journal} {\bibinfo  {journal}
  {{NEW JOURNAL OF PHYSICS}}\ }\textbf {\bibinfo {volume} {{10}}} (\bibinfo
  {year} {{2008}}),\ {10.1088/1367-2630/10/12/125027}}\BibitemShut {NoStop}%
\bibitem [{\citenamefont {Dai}\ \emph {et~al.}(2014)\citenamefont {Dai},
  \citenamefont {Calleja}, \citenamefont {Alldredge}, \citenamefont {Zhu},
  \citenamefont {Li}, \citenamefont {Lu}, \citenamefont {Sun}, \citenamefont
  {Wolf}, \citenamefont {Berger},\ and\ \citenamefont {McElroy}}]{Dai2014}%
  \BibitemOpen
  \bibfield  {author} {\bibinfo {author} {\bibfnamefont {J.}~\bibnamefont
  {Dai}}, \bibinfo {author} {\bibfnamefont {E.}~\bibnamefont {Calleja}},
  \bibinfo {author} {\bibfnamefont {J.}~\bibnamefont {Alldredge}}, \bibinfo
  {author} {\bibfnamefont {X.}~\bibnamefont {Zhu}}, \bibinfo {author}
  {\bibfnamefont {L.}~\bibnamefont {Li}}, \bibinfo {author} {\bibfnamefont
  {W.}~\bibnamefont {Lu}}, \bibinfo {author} {\bibfnamefont {Y.}~\bibnamefont
  {Sun}}, \bibinfo {author} {\bibfnamefont {T.}~\bibnamefont {Wolf}}, \bibinfo
  {author} {\bibfnamefont {H.}~\bibnamefont {Berger}}, \ and\ \bibinfo {author}
  {\bibfnamefont {K.}~\bibnamefont {McElroy}},\ }\href {\doibase
  10.1103/PhysRevB.89.165140} {\bibfield  {journal} {\bibinfo  {journal} {Phys.
  Rev. B}\ }\textbf {\bibinfo {volume} {89}},\ \bibinfo {pages} {165140}
  (\bibinfo {year} {2014})}\BibitemShut {NoStop}%
\bibitem [{\citenamefont {Fang}\ \emph {et~al.}(2013)\citenamefont {Fang},
  \citenamefont {Xu}, \citenamefont {Dong}, \citenamefont {Zheng},\ and\
  \citenamefont {Wang}}]{Fang2013}%
  \BibitemOpen
  \bibfield  {author} {\bibinfo {author} {\bibfnamefont {A.}~\bibnamefont
  {Fang}}, \bibinfo {author} {\bibfnamefont {G.}~\bibnamefont {Xu}}, \bibinfo
  {author} {\bibfnamefont {T.}~\bibnamefont {Dong}}, \bibinfo {author}
  {\bibfnamefont {P.}~\bibnamefont {Zheng}}, \ and\ \bibinfo {author}
  {\bibfnamefont {N.}~\bibnamefont {Wang}},\ }\href@noop {} {\bibfield
  {journal} {\bibinfo  {journal} {Sci. Rep.}\ }\textbf {\bibinfo {volume}
  {3}},\ \bibinfo {pages} {1153} (\bibinfo {year} {2013})}\BibitemShut
  {NoStop}%
\bibitem [{\citenamefont {Ootsuki}\ \emph {et~al.}(2012)\citenamefont
  {Ootsuki}, \citenamefont {Wakisaka}, \citenamefont {Pyon}, \citenamefont
  {Kudo}, \citenamefont {Nohara}, \citenamefont {Arita}, \citenamefont
  {Namatame}, \citenamefont {Taniguchi}, \citenamefont {Saini},\ and\
  \citenamefont {Mizokawa}}]{Ootsuki2012a}%
  \BibitemOpen
  \bibfield  {author} {\bibinfo {author} {\bibfnamefont {D.}~\bibnamefont
  {Ootsuki}}, \bibinfo {author} {\bibfnamefont {Y.}~\bibnamefont {Wakisaka}},
  \bibinfo {author} {\bibfnamefont {S.}~\bibnamefont {Pyon}}, \bibinfo {author}
  {\bibfnamefont {K.}~\bibnamefont {Kudo}}, \bibinfo {author} {\bibfnamefont
  {M.}~\bibnamefont {Nohara}}, \bibinfo {author} {\bibfnamefont
  {M.}~\bibnamefont {Arita}}, \bibinfo {author} {\bibfnamefont
  {H.}~\bibnamefont {Namatame}}, \bibinfo {author} {\bibfnamefont
  {M.}~\bibnamefont {Taniguchi}}, \bibinfo {author} {\bibfnamefont
  {N.}~\bibnamefont {Saini}}, \ and\ \bibinfo {author} {\bibfnamefont
  {T.}~\bibnamefont {Mizokawa}},\ }\href@noop {} {\bibfield  {journal}
  {\bibinfo  {journal} {Phys. Rev. B}\ }\textbf {\bibinfo {volume} {86}},\
  \bibinfo {pages} {014519} (\bibinfo {year} {2012})}\BibitemShut {NoStop}%
\bibitem [{\citenamefont {Pyon}\ \emph {et~al.}(2012)\citenamefont {Pyon},
  \citenamefont {Kudo},\ and\ \citenamefont {Nohara}}]{Pyon2012}%
  \BibitemOpen
  \bibfield  {author} {\bibinfo {author} {\bibfnamefont {S.}~\bibnamefont
  {Pyon}}, \bibinfo {author} {\bibfnamefont {K.}~\bibnamefont {Kudo}}, \ and\
  \bibinfo {author} {\bibfnamefont {M.}~\bibnamefont {Nohara}},\ }\href@noop {}
  {\bibfield  {journal} {\bibinfo  {journal} {J. Phys. Soc. Jpn.}\ }\textbf
  {\bibinfo {volume} {81}},\ \bibinfo {pages} {053701} (\bibinfo {year}
  {2012})}\BibitemShut {NoStop}%
\bibitem [{\citenamefont {Yang}\ \emph {et~al.}(2012)\citenamefont {Yang},
  \citenamefont {Choi}, \citenamefont {Oh}, \citenamefont {Hogan},
  \citenamefont {Horibe}, \citenamefont {Kim}, \citenamefont {Min},\ and\
  \citenamefont {Cheong}}]{Yang2012}%
  \BibitemOpen
  \bibfield  {author} {\bibinfo {author} {\bibfnamefont {J.~J.}\ \bibnamefont
  {Yang}}, \bibinfo {author} {\bibfnamefont {Y.~J.}\ \bibnamefont {Choi}},
  \bibinfo {author} {\bibfnamefont {Y.~S.}\ \bibnamefont {Oh}}, \bibinfo
  {author} {\bibfnamefont {A.}~\bibnamefont {Hogan}}, \bibinfo {author}
  {\bibfnamefont {Y.}~\bibnamefont {Horibe}}, \bibinfo {author} {\bibfnamefont
  {K.}~\bibnamefont {Kim}}, \bibinfo {author} {\bibfnamefont {B.~I.}\
  \bibnamefont {Min}}, \ and\ \bibinfo {author} {\bibfnamefont {S.~W.}\
  \bibnamefont {Cheong}},\ }\href@noop {} {\bibfield  {journal} {\bibinfo
  {journal} {Phys. Rev. Lett.}\ }\textbf {\bibinfo {volume} {108}},\ \bibinfo
  {pages} {116402} (\bibinfo {year} {2012})}\BibitemShut {NoStop}%
\bibitem [{\citenamefont {Kudo}\ \emph {et~al.}(2013)\citenamefont {Kudo},
  \citenamefont {Kobayashi}, \citenamefont {Pyon},\ and\ \citenamefont
  {Nohara}}]{Kudo2013}%
  \BibitemOpen
  \bibfield  {author} {\bibinfo {author} {\bibfnamefont {K.}~\bibnamefont
  {Kudo}}, \bibinfo {author} {\bibfnamefont {M.}~\bibnamefont {Kobayashi}},
  \bibinfo {author} {\bibfnamefont {S.}~\bibnamefont {Pyon}}, \ and\ \bibinfo
  {author} {\bibfnamefont {M.}~\bibnamefont {Nohara}},\ }\href@noop {}
  {\bibfield  {journal} {\bibinfo  {journal} {J. Phys. Soc. Jpn.}\ }\textbf
  {\bibinfo {volume} {82}},\ \bibinfo {pages} {085001} (\bibinfo {year}
  {2013})}\BibitemShut {NoStop}%
\bibitem [{\citenamefont {Matsumoto}\ \emph {et~al.}(1999)\citenamefont
  {Matsumoto}, \citenamefont {Kouji}, \citenamefont {Endoh}, \citenamefont
  {Takano},\ and\ \citenamefont {Nagata}}]{Matsumoto1999}%
  \BibitemOpen
  \bibfield  {author} {\bibinfo {author} {\bibfnamefont {N.}~\bibnamefont
  {Matsumoto}}, \bibinfo {author} {\bibfnamefont {T.}~\bibnamefont {Kouji}},
  \bibinfo {author} {\bibfnamefont {R.}~\bibnamefont {Endoh}}, \bibinfo
  {author} {\bibfnamefont {H.}~\bibnamefont {Takano}}, \ and\ \bibinfo {author}
  {\bibfnamefont {S.}~\bibnamefont {Nagata}},\ }\href@noop {} {\bibfield
  {journal} {\bibinfo  {journal} {J. Low Temp. Phys.}\ }\textbf {\bibinfo
  {volume} {117}},\ \bibinfo {pages} {1129} (\bibinfo {year}
  {1999})}\BibitemShut {NoStop}%
\bibitem [{\citenamefont {Oh}\ \emph {et~al.}(2013)\citenamefont {Oh},
  \citenamefont {Yang}, \citenamefont {Horibe},\ and\ \citenamefont
  {Cheong}}]{Oh2013}%
  \BibitemOpen
  \bibfield  {author} {\bibinfo {author} {\bibfnamefont {Y.~S.}\ \bibnamefont
  {Oh}}, \bibinfo {author} {\bibfnamefont {J.~J.}\ \bibnamefont {Yang}},
  \bibinfo {author} {\bibfnamefont {Y.}~\bibnamefont {Horibe}}, \ and\ \bibinfo
  {author} {\bibfnamefont {S.~W.}\ \bibnamefont {Cheong}},\ }\href@noop {}
  {\bibfield  {journal} {\bibinfo  {journal} {Phys. Rev. Lett.}\ }\textbf
  {\bibinfo {volume} {110}},\ \bibinfo {pages} {127209} (\bibinfo {year}
  {2013})}\BibitemShut {NoStop}%
\bibitem [{\citenamefont {Qian}\ \emph {et~al.}(2013)\citenamefont {Qian},
  \citenamefont {Miao}, \citenamefont {Wang}, \citenamefont {Liu},
  \citenamefont {Shi}, \citenamefont {Huang}, \citenamefont {Zhang},
  \citenamefont {Xu}, \citenamefont {Richard}, \citenamefont {Shi},
  \citenamefont {Upton}, \citenamefont {Hill}, \citenamefont {Xu},
  \citenamefont {Dai}, \citenamefont {Fang}, \citenamefont {Lei}, \citenamefont
  {Petrovic}, \citenamefont {Fang}, \citenamefont {Wang},\ and\ \citenamefont
  {Ding}}]{Qian2013}%
  \BibitemOpen
  \bibfield  {author} {\bibinfo {author} {\bibfnamefont {T.}~\bibnamefont
  {Qian}}, \bibinfo {author} {\bibfnamefont {H.}~\bibnamefont {Miao}}, \bibinfo
  {author} {\bibfnamefont {Z.~J.}\ \bibnamefont {Wang}}, \bibinfo {author}
  {\bibfnamefont {X.}~\bibnamefont {Liu}}, \bibinfo {author} {\bibfnamefont
  {X.}~\bibnamefont {Shi}}, \bibinfo {author} {\bibfnamefont {Y.~B.}\
  \bibnamefont {Huang}}, \bibinfo {author} {\bibfnamefont {P.}~\bibnamefont
  {Zhang}}, \bibinfo {author} {\bibfnamefont {N.}~\bibnamefont {Xu}}, \bibinfo
  {author} {\bibfnamefont {P.}~\bibnamefont {Richard}}, \bibinfo {author}
  {\bibfnamefont {M.}~\bibnamefont {Shi}}, \bibinfo {author} {\bibfnamefont
  {M.~H.}\ \bibnamefont {Upton}}, \bibinfo {author} {\bibfnamefont {J.~P.}\
  \bibnamefont {Hill}}, \bibinfo {author} {\bibfnamefont {G.}~\bibnamefont
  {Xu}}, \bibinfo {author} {\bibfnamefont {X.}~\bibnamefont {Dai}}, \bibinfo
  {author} {\bibfnamefont {Z.}~\bibnamefont {Fang}}, \bibinfo {author}
  {\bibfnamefont {H.~C.}\ \bibnamefont {Lei}}, \bibinfo {author} {\bibfnamefont
  {C.}~\bibnamefont {Petrovic}}, \bibinfo {author} {\bibfnamefont {A.~F.}\
  \bibnamefont {Fang}}, \bibinfo {author} {\bibfnamefont {N.~L.}\ \bibnamefont
  {Wang}}, \ and\ \bibinfo {author} {\bibfnamefont {H.}~\bibnamefont {Ding}},\
  }\href@noop {} {\  (\bibinfo {year} {2013})},\ \Eprint
  {http://arxiv.org/abs/1311.4946} {arXiv:1311.4946} \BibitemShut {NoStop}%
\bibitem [{\citenamefont {Ootsuki}\ \emph {et~al.}(2013)\citenamefont
  {Ootsuki}, \citenamefont {Pyon}, \citenamefont {Kudo}, \citenamefont
  {Nohara}, \citenamefont {Horio}, \citenamefont {Yoshida}, \citenamefont
  {Fujimori}, \citenamefont {Arita}, \citenamefont {Anzai}, \citenamefont
  {Namatame}, \citenamefont {Taniguchi}, \citenamefont {Saini},\ and\
  \citenamefont {Mizokawa}}]{Ootsuki2012}%
  \BibitemOpen
  \bibfield  {author} {\bibinfo {author} {\bibfnamefont {D.}~\bibnamefont
  {Ootsuki}}, \bibinfo {author} {\bibfnamefont {S.}~\bibnamefont {Pyon}},
  \bibinfo {author} {\bibfnamefont {K.}~\bibnamefont {Kudo}}, \bibinfo {author}
  {\bibfnamefont {M.}~\bibnamefont {Nohara}}, \bibinfo {author} {\bibfnamefont
  {M.}~\bibnamefont {Horio}}, \bibinfo {author} {\bibfnamefont
  {T.}~\bibnamefont {Yoshida}}, \bibinfo {author} {\bibfnamefont
  {A.}~\bibnamefont {Fujimori}}, \bibinfo {author} {\bibfnamefont
  {M.}~\bibnamefont {Arita}}, \bibinfo {author} {\bibfnamefont
  {H.}~\bibnamefont {Anzai}}, \bibinfo {author} {\bibfnamefont
  {H.}~\bibnamefont {Namatame}}, \bibinfo {author} {\bibfnamefont
  {M.}~\bibnamefont {Taniguchi}}, \bibinfo {author} {\bibfnamefont
  {N.}~\bibnamefont {Saini}}, \ and\ \bibinfo {author} {\bibfnamefont
  {T.}~\bibnamefont {Mizokawa}},\ }\href@noop {} {\bibfield  {journal}
  {\bibinfo  {journal} {J. Phys. Soc. Jpn.}\ }\textbf {\bibinfo {volume}
  {82}},\ \bibinfo {pages} {093704} (\bibinfo {year} {2013})},\ \Eprint
  {http://arxiv.org/abs/1207.2613v1} {arXiv:1207.2613v1} \BibitemShut {NoStop}%
\bibitem [{\citenamefont {Mizuno}\ \emph {et~al.}(2002)\citenamefont {Mizuno},
  \citenamefont {Magishi}, \citenamefont {Shinonome}, \citenamefont {Saito},
  \citenamefont {Koyama}, \citenamefont {Matsumoto},\ and\ \citenamefont
  {Nagata}}]{Mizuno2002}%
  \BibitemOpen
  \bibfield  {author} {\bibinfo {author} {\bibfnamefont {K.}~\bibnamefont
  {Mizuno}}, \bibinfo {author} {\bibfnamefont {K.}~\bibnamefont {Magishi}},
  \bibinfo {author} {\bibfnamefont {Y.}~\bibnamefont {Shinonome}}, \bibinfo
  {author} {\bibfnamefont {T.}~\bibnamefont {Saito}}, \bibinfo {author}
  {\bibfnamefont {K.}~\bibnamefont {Koyama}}, \bibinfo {author} {\bibfnamefont
  {N.}~\bibnamefont {Matsumoto}}, \ and\ \bibinfo {author} {\bibfnamefont
  {S.}~\bibnamefont {Nagata}},\ }\href@noop {} {\bibfield  {journal} {\bibinfo
  {journal} {Physica B}\ }\textbf {\bibinfo {volume} {312}},\ \bibinfo {pages}
  {818} (\bibinfo {year} {2002})}\BibitemShut {NoStop}%
\bibitem [{\citenamefont {Kiswandhi}\ \emph {et~al.}(2013)\citenamefont
  {Kiswandhi}, \citenamefont {Brooks}, \citenamefont {Cao}, \citenamefont
  {Yan}, \citenamefont {Mandrus}, \citenamefont {Jiang},\ and\ \citenamefont
  {Zhou}}]{Kiswandhi2013}%
  \BibitemOpen
  \bibfield  {author} {\bibinfo {author} {\bibfnamefont {A.}~\bibnamefont
  {Kiswandhi}}, \bibinfo {author} {\bibfnamefont {J.~S.}\ \bibnamefont
  {Brooks}}, \bibinfo {author} {\bibfnamefont {H.~B.}\ \bibnamefont {Cao}},
  \bibinfo {author} {\bibfnamefont {J.~Q.}\ \bibnamefont {Yan}}, \bibinfo
  {author} {\bibfnamefont {D.}~\bibnamefont {Mandrus}}, \bibinfo {author}
  {\bibfnamefont {Z.}~\bibnamefont {Jiang}}, \ and\ \bibinfo {author}
  {\bibfnamefont {H.~D.}\ \bibnamefont {Zhou}},\ }\href@noop {} {\bibfield
  {journal} {\bibinfo  {journal} {Phys. Rev. B}\ }\textbf {\bibinfo {volume}
  {87}},\ \bibinfo {pages} {121107} (\bibinfo {year} {2013})}\BibitemShut
  {NoStop}%
\bibitem [{\citenamefont {Pascut}\ \emph {et~al.}(2014)\citenamefont {Pascut},
  \citenamefont {Haule}, \citenamefont {Gutmann}, \citenamefont {Barnett},
  \citenamefont {Bombardi}, \citenamefont {Artyukhin}, \citenamefont
  {Vanderbilt}, \citenamefont {Yang}, \citenamefont {Cheong},\ and\
  \citenamefont {Kiryukhin}}]{Pascut}%
  \BibitemOpen
  \bibfield  {author} {\bibinfo {author} {\bibfnamefont {G.~L.}\ \bibnamefont
  {Pascut}}, \bibinfo {author} {\bibfnamefont {K.}~\bibnamefont {Haule}},
  \bibinfo {author} {\bibfnamefont {M.~J.}\ \bibnamefont {Gutmann}}, \bibinfo
  {author} {\bibfnamefont {S.~A.}\ \bibnamefont {Barnett}}, \bibinfo {author}
  {\bibfnamefont {A.}~\bibnamefont {Bombardi}}, \bibinfo {author}
  {\bibfnamefont {S.}~\bibnamefont {Artyukhin}}, \bibinfo {author}
  {\bibfnamefont {D.}~\bibnamefont {Vanderbilt}}, \bibinfo {author}
  {\bibfnamefont {J.~J.}\ \bibnamefont {Yang}}, \bibinfo {author}
  {\bibfnamefont {S.~W.}\ \bibnamefont {Cheong}}, \ and\ \bibinfo {author}
  {\bibfnamefont {V.}~\bibnamefont {Kiryukhin}},\ }\href@noop {} {\bibfield
  {journal} {\bibinfo  {journal} {Phys. Rev. Lett.}\ }\textbf {\bibinfo
  {volume} {112}},\ \bibinfo {pages} {086402} (\bibinfo {year} {2014})},\
  \Eprint {http://arxiv.org/abs/1309.3548} {1309.3548} \BibitemShut {NoStop}%
\bibitem [{\citenamefont {Cao}\ \emph {et~al.}(2013)\citenamefont {Cao},
  \citenamefont {Chakoumakos}, \citenamefont {Chen}, \citenamefont {Yan},
  \citenamefont {Mcguire}, \citenamefont {Yang}, \citenamefont {Custelcean},
  \citenamefont {Zhou}, \citenamefont {Singh},\ and\ \citenamefont
  {Mandrus}}]{Cao2013a}%
  \BibitemOpen
  \bibfield  {author} {\bibinfo {author} {\bibfnamefont {H.}~\bibnamefont
  {Cao}}, \bibinfo {author} {\bibfnamefont {B.~C.}\ \bibnamefont
  {Chakoumakos}}, \bibinfo {author} {\bibfnamefont {X.}~\bibnamefont {Chen}},
  \bibinfo {author} {\bibfnamefont {J.}~\bibnamefont {Yan}}, \bibinfo {author}
  {\bibfnamefont {M.~A.}\ \bibnamefont {Mcguire}}, \bibinfo {author}
  {\bibfnamefont {H.}~\bibnamefont {Yang}}, \bibinfo {author} {\bibfnamefont
  {R.}~\bibnamefont {Custelcean}}, \bibinfo {author} {\bibfnamefont
  {H.}~\bibnamefont {Zhou}}, \bibinfo {author} {\bibfnamefont {D.~J.}\
  \bibnamefont {Singh}}, \ and\ \bibinfo {author} {\bibfnamefont
  {D.}~\bibnamefont {Mandrus}},\ }\href@noop {} {\bibfield  {journal} {\bibinfo
   {journal} {Phys. Rev. B}\ }\textbf {\bibinfo {volume} {88}},\ \bibinfo
  {pages} {115122} (\bibinfo {year} {2013})}\BibitemShut {NoStop}%
\bibitem [{\citenamefont {Hsu}\ \emph {et~al.}(2013)\citenamefont {Hsu},
  \citenamefont {Mauerer}, \citenamefont {Vogt}, \citenamefont {Yang},
  \citenamefont {Oh}, \citenamefont {Cheong}, \citenamefont {Bode},\ and\
  \citenamefont {Wu}}]{Hsu2013}%
  \BibitemOpen
  \bibfield  {author} {\bibinfo {author} {\bibfnamefont {P.~J.}\ \bibnamefont
  {Hsu}}, \bibinfo {author} {\bibfnamefont {T.}~\bibnamefont {Mauerer}},
  \bibinfo {author} {\bibfnamefont {M.}~\bibnamefont {Vogt}}, \bibinfo {author}
  {\bibfnamefont {J.~J.}\ \bibnamefont {Yang}}, \bibinfo {author}
  {\bibfnamefont {S.}~\bibnamefont {Oh}}, \bibinfo {author} {\bibfnamefont
  {S.~W.}\ \bibnamefont {Cheong}}, \bibinfo {author} {\bibfnamefont
  {M.}~\bibnamefont {Bode}}, \ and\ \bibinfo {author} {\bibfnamefont
  {W.}~\bibnamefont {Wu}},\ }\href@noop {} {\bibfield  {journal} {\bibinfo
  {journal} {Phys. Rev. Lett.}\ }\textbf {\bibinfo {volume} {111}},\ \bibinfo
  {pages} {266401} (\bibinfo {year} {2013})}\BibitemShut {NoStop}%
\bibitem [{\citenamefont {Toriyama}\ \emph {et~al.}(2014)\citenamefont
  {Toriyama}, \citenamefont {Kobori}, \citenamefont {Konishi}, \citenamefont
  {Ohta}, \citenamefont {Sugimoto}, \citenamefont {Kim}, \citenamefont
  {Fujiwara}, \citenamefont {Pyon}, \citenamefont {Kudo},\ and\ \citenamefont
  {Nohara}}]{Toriyama2014}%
  \BibitemOpen
  \bibfield  {author} {\bibinfo {author} {\bibfnamefont {T.}~\bibnamefont
  {Toriyama}}, \bibinfo {author} {\bibfnamefont {M.}~\bibnamefont {Kobori}},
  \bibinfo {author} {\bibfnamefont {T.}~\bibnamefont {Konishi}}, \bibinfo
  {author} {\bibfnamefont {Y.}~\bibnamefont {Ohta}}, \bibinfo {author}
  {\bibfnamefont {K.}~\bibnamefont {Sugimoto}}, \bibinfo {author}
  {\bibfnamefont {J.}~\bibnamefont {Kim}}, \bibinfo {author} {\bibfnamefont
  {A.}~\bibnamefont {Fujiwara}}, \bibinfo {author} {\bibfnamefont
  {S.}~\bibnamefont {Pyon}}, \bibinfo {author} {\bibfnamefont {K.}~\bibnamefont
  {Kudo}}, \ and\ \bibinfo {author} {\bibfnamefont {M.}~\bibnamefont
  {Nohara}},\ }\href@noop {} {\bibfield  {journal} {\bibinfo  {journal}
  {Journal of the Physical Society of Japan}\ }\textbf {\bibinfo {volume}
  {83}},\ \bibinfo {pages} {033701} (\bibinfo {year} {2014})}\BibitemShut
  {NoStop}%
\bibitem [{\citenamefont {Ootsuki}\ \emph
  {et~al.}(2014{\natexlab{a}})\citenamefont {Ootsuki}, \citenamefont
  {Toriyama}, \citenamefont {Kobayashi}, \citenamefont {Pyon}, \citenamefont
  {Kudo}, \citenamefont {Nohara}, \citenamefont {Sugimoto}, \citenamefont
  {Yoshida}, \citenamefont {Horio}, \citenamefont {Fujimori}, \citenamefont
  {Arita}, \citenamefont {Anzai}, \citenamefont {Namatame}, \citenamefont
  {Taniguchi}, \citenamefont {Saini}, \citenamefont {Konishi}, \citenamefont
  {Ohta},\ and\ \citenamefont {Mizokawa}}]{Ootsuki2013}%
  \BibitemOpen
  \bibfield  {author} {\bibinfo {author} {\bibfnamefont {D.}~\bibnamefont
  {Ootsuki}}, \bibinfo {author} {\bibfnamefont {T.}~\bibnamefont {Toriyama}},
  \bibinfo {author} {\bibfnamefont {M.}~\bibnamefont {Kobayashi}}, \bibinfo
  {author} {\bibfnamefont {S.}~\bibnamefont {Pyon}}, \bibinfo {author}
  {\bibfnamefont {K.}~\bibnamefont {Kudo}}, \bibinfo {author} {\bibfnamefont
  {M.}~\bibnamefont {Nohara}}, \bibinfo {author} {\bibfnamefont
  {T.}~\bibnamefont {Sugimoto}}, \bibinfo {author} {\bibfnamefont
  {T.}~\bibnamefont {Yoshida}}, \bibinfo {author} {\bibfnamefont
  {M.}~\bibnamefont {Horio}}, \bibinfo {author} {\bibfnamefont
  {A.}~\bibnamefont {Fujimori}}, \bibinfo {author} {\bibfnamefont
  {M.}~\bibnamefont {Arita}}, \bibinfo {author} {\bibfnamefont
  {H.}~\bibnamefont {Anzai}}, \bibinfo {author} {\bibfnamefont
  {H.}~\bibnamefont {Namatame}}, \bibinfo {author} {\bibfnamefont
  {M.}~\bibnamefont {Taniguchi}}, \bibinfo {author} {\bibfnamefont {N.~L.}\
  \bibnamefont {Saini}}, \bibinfo {author} {\bibfnamefont {T.}~\bibnamefont
  {Konishi}}, \bibinfo {author} {\bibfnamefont {Y.}~\bibnamefont {Ohta}}, \
  and\ \bibinfo {author} {\bibfnamefont {T.}~\bibnamefont {Mizokawa}},\
  }\href@noop {} {\bibfield  {journal} {\bibinfo  {journal} {J. Phys. Soc.
  Jpn.}\ }\textbf {\bibinfo {volume} {83}},\ \bibinfo {pages} {033704}
  (\bibinfo {year} {2014}{\natexlab{a}})},\ \Eprint
  {http://arxiv.org/abs/1311.1199} {1311.1199} \BibitemShut {NoStop}%
\bibitem [{\citenamefont {Ootsuki}\ \emph
  {et~al.}(2014{\natexlab{b}})\citenamefont {Ootsuki}, \citenamefont
  {Toriyama}, \citenamefont {Kobayashi}, \citenamefont {Pyon}, \citenamefont
  {Kudo}, \citenamefont {Nohara}, \citenamefont {Horiba}, \citenamefont {Ono},
  \citenamefont {Kumigashira}, \citenamefont {Noda}, \citenamefont {Sugimoto},
  \citenamefont {Fujimori}, \citenamefont {Saini}, \citenamefont {Konishi},
  \citenamefont {Ohta},\ and\ \citenamefont {Mizokawa}}]{Ootsuki2013a}%
  \BibitemOpen
  \bibfield  {author} {\bibinfo {author} {\bibfnamefont {D.}~\bibnamefont
  {Ootsuki}}, \bibinfo {author} {\bibfnamefont {T.}~\bibnamefont {Toriyama}},
  \bibinfo {author} {\bibfnamefont {M.}~\bibnamefont {Kobayashi}}, \bibinfo
  {author} {\bibfnamefont {S.}~\bibnamefont {Pyon}}, \bibinfo {author}
  {\bibfnamefont {K.}~\bibnamefont {Kudo}}, \bibinfo {author} {\bibfnamefont
  {M.}~\bibnamefont {Nohara}}, \bibinfo {author} {\bibfnamefont
  {K.}~\bibnamefont {Horiba}}, \bibinfo {author} {\bibfnamefont
  {K.}~\bibnamefont {Ono}}, \bibinfo {author} {\bibfnamefont {H.}~\bibnamefont
  {Kumigashira}}, \bibinfo {author} {\bibfnamefont {T.}~\bibnamefont {Noda}},
  \bibinfo {author} {\bibfnamefont {T.}~\bibnamefont {Sugimoto}}, \bibinfo
  {author} {\bibfnamefont {A.}~\bibnamefont {Fujimori}}, \bibinfo {author}
  {\bibfnamefont {N.~L.}\ \bibnamefont {Saini}}, \bibinfo {author}
  {\bibfnamefont {T.}~\bibnamefont {Konishi}}, \bibinfo {author} {\bibfnamefont
  {Y.}~\bibnamefont {Ohta}}, \ and\ \bibinfo {author} {\bibfnamefont
  {T.}~\bibnamefont {Mizokawa}},\ }\href@noop {} {\bibfield  {journal}
  {\bibinfo  {journal} {Phys. Rev. B}\ }\textbf {\bibinfo {volume} {89}},\
  \bibinfo {pages} {104506} (\bibinfo {year} {2014}{\natexlab{b}})}\BibitemShut
  {NoStop}%
\bibitem [{\citenamefont {Pyon}\ \emph {et~al.}(2013)\citenamefont {Pyon},
  \citenamefont {Kudo},\ and\ \citenamefont {Nohara}}]{Pyon2013}%
  \BibitemOpen
  \bibfield  {author} {\bibinfo {author} {\bibfnamefont {S.}~\bibnamefont
  {Pyon}}, \bibinfo {author} {\bibfnamefont {K.}~\bibnamefont {Kudo}}, \ and\
  \bibinfo {author} {\bibfnamefont {M.}~\bibnamefont {Nohara}},\ }\href@noop {}
  {\bibfield  {journal} {\bibinfo  {journal} {Physica C: Superconductivity}\
  }\textbf {\bibinfo {volume} {494}},\ \bibinfo {pages} {80} (\bibinfo {year}
  {2013})}\BibitemShut {NoStop}%
\bibitem [{\citenamefont {Blaha}\ \emph {et~al.}(2001)\citenamefont {Blaha},
  \citenamefont {Schwarz}, \citenamefont {Madsen}, \citenamefont {Kvasnicka},\
  and\ \citenamefont {Luitz}}]{Blaha2001}%
  \BibitemOpen
  \bibfield  {author} {\bibinfo {author} {\bibfnamefont {P.}~\bibnamefont
  {Blaha}}, \bibinfo {author} {\bibfnamefont {K.}~\bibnamefont {Schwarz}},
  \bibinfo {author} {\bibfnamefont {G.}~\bibnamefont {Madsen}}, \bibinfo
  {author} {\bibfnamefont {D.}~\bibnamefont {Kvasnicka}}, \ and\ \bibinfo
  {author} {\bibfnamefont {J.}~\bibnamefont {Luitz}},\ }\href@noop {} {\enquote
  {\bibinfo {title} {{WIEN2k, An Augmented Plane Wave + Local Orbitals Program
  for Calculating Crystal Properties}},}\ } (\bibinfo {year}
  {2001})\BibitemShut {NoStop}%
\bibitem [{Note1()}]{Note1}%
  \BibitemOpen
  \bibinfo {note} {The cutoff for plane wave expansion in the interstitials was
  $R \times K_{max}=9$ (convergence is reached for $R \times K_{max}=8$) and
  5472 k-points were used in the first Brillouin zone.}\BibitemShut {Stop}%
\bibitem [{\citenamefont {Rourke}\ and\ \citenamefont
  {Julian}(2012)}]{Rourke2012}%
  \BibitemOpen
  \bibfield  {author} {\bibinfo {author} {\bibfnamefont {P.}~\bibnamefont
  {Rourke}}\ and\ \bibinfo {author} {\bibfnamefont {S.}~\bibnamefont
  {Julian}},\ }\href@noop {} {\bibfield  {journal} {\bibinfo  {journal} {Comp.
  Phys. Comm.}\ }\textbf {\bibinfo {volume} {183}},\ \bibinfo {pages} {324}
  (\bibinfo {year} {2012})}\BibitemShut {NoStop}%
\bibitem [{\citenamefont {Steep}\ \emph {et~al.}(1995)\citenamefont {Steep},
  \citenamefont {Rettenberger}, \citenamefont {Meyer}, \citenamefont {Jansen},
  \citenamefont {Joss}, \citenamefont {Biberacher}, \citenamefont {Bucher},\
  and\ \citenamefont {Oglesby}}]{Steep1995}%
  \BibitemOpen
  \bibfield  {author} {\bibinfo {author} {\bibfnamefont {E.}~\bibnamefont
  {Steep}}, \bibinfo {author} {\bibfnamefont {S.}~\bibnamefont {Rettenberger}},
  \bibinfo {author} {\bibfnamefont {F.}~\bibnamefont {Meyer}}, \bibinfo
  {author} {\bibfnamefont {A.}~\bibnamefont {Jansen}}, \bibinfo {author}
  {\bibfnamefont {W.}~\bibnamefont {Joss}}, \bibinfo {author} {\bibfnamefont
  {W.}~\bibnamefont {Biberacher}}, \bibinfo {author} {\bibfnamefont
  {E.}~\bibnamefont {Bucher}}, \ and\ \bibinfo {author} {\bibfnamefont
  {C.}~\bibnamefont {Oglesby}},\ }\href {\doibase
  http://dx.doi.org/10.1016/0921-4526(94)00258-W} {\bibfield  {journal}
  {\bibinfo  {journal} {Physica B: Condensed Matter}\ }\textbf {\bibinfo
  {volume} {204}},\ \bibinfo {pages} {162 } (\bibinfo {year}
  {1995})}\BibitemShut {NoStop}%
\bibitem [{\citenamefont {Lifshitz}\ and\ \citenamefont
  {Kosevich}(1956)}]{Lifshitz1956}%
  \BibitemOpen
  \bibfield  {author} {\bibinfo {author} {\bibfnamefont {L.}~\bibnamefont
  {Lifshitz}}\ and\ \bibinfo {author} {\bibfnamefont {A.}~\bibnamefont
  {Kosevich}},\ }\href@noop {} {\bibfield  {journal} {\bibinfo  {journal} {Sov.
  Phys. JETP}\ }\textbf {\bibinfo {volume} {2}},\ \bibinfo {pages} {636}
  (\bibinfo {year} {1956})}\BibitemShut {NoStop}%
\bibitem [{Note2()}]{Note2}%
  \BibitemOpen
  \bibinfo {note} {For large $F$ and $m^*$, the exponential damping of the dHvA
  signal due impurity scattering will hamper the observation of those dHvA
  oscillations, as compared with small $F$ and $m^*$. Any small misalignment of
  the samples could cause small shifts in the positions of the dHvA
  frequencies}\BibitemShut {NoStop}%
\end{thebibliography}%

\end{document}